\documentclass[12pt]{article}   
\usepackage{epsfig}

\newlength{\dinwidth} \newlength{\dinmargin}
\begin{document}
\begin {flushright}
Cavendish-HEP-04/20\\
FSU-HEP-040702
\end {flushright} 
\vspace{3mm}

\begin{center}
{\Large \bf Anomalous $tqV$ couplings and FCNC \\top quark 
production\footnote{Presented
at the DIS 2004 Workshop, Strbske Pleso, Slovakia, 14-18 April, 2004.}}
\end{center}
\vspace{2mm}

\begin{center}
{\large Nikolaos Kidonakis$^a$ and Alexander Belyaev$^b$}\\
\vspace{2mm}
{\it $^a$Cavendish Laboratory, University of Cambridge\\
Madingley Road, Cambridge CB3 0HE, UK\\
E-mail: kidonaki@hep.phy.cam.ac.uk\\
\vspace{1mm}
$^b$Physics Department, Florida State University\\
Tallahassee, FL 32306-4350, USA\\
E-mail: belyaev@hep.fsu.edu}
\end{center}

\vspace{3mm}

\begin{abstract}

We discuss FCNC top quark production via anomalous
$tqV$ couplings at the Tevatron and HERA colliders.
We calculate higher-order soft-gluon corrections to such processes 
and demonstrate the stabilization of the cross section when these 
corrections are included. 

\end{abstract}

\thispagestyle{empty} \newpage \setcounter{page}{2}

\section{Introduction} 

Flavor-changing neutral-current (FCNC) processes involving the top quark
appear in several models of physics beyond the Standard Model. 
The effective Lagrangian involving anomalous $tqV$ couplings can be written as
$
\Delta {\cal L}^{eff} =    \frac{1}{ \Lambda } \,
\kappa_{tqV} \, e \, \bar t \, \sigma_{\mu\nu} \, q \, F^{\mu\nu}_V + h.c.
$
where  $\kappa_{tqV}$ is the anomalous coupling, with $q$ denoting an 
up or charm quark and $V$ a photon or $Z$-boson with field tensor 
$F_V^{\mu\nu}$;
$\sigma_{\mu \nu}=(i/2)(\gamma_{\mu}\gamma_{\nu}
-\gamma_{\nu}\gamma_{\mu})$ with $\gamma^{\mu}$ the Dirac matrices;
and $\Lambda$ is an effective scale which we  set equal to the top quark
mass, $m$.

The present TeV energy scale colliders --
Tevatron and HERA-- can probe FCNC interactions in the top-quark sector
and set limits on $\kappa_{tq\gamma}$ and $\kappa_{tqZ}$.  
However, there are large uncertainties in the lowest-order results from 
variation of the factorization/renormalization scales, $\mu$.
Therefore  the stabilization of the cross section for these FCNC
processes is timely and important.
We have calculated next-to-leading order (NLO) and
next-to-next-to-leading order (NNLO) soft-gluon corrections
for the following processes: $gu \rightarrow tZ$,
$gu \rightarrow t\gamma$, and 
$uu \rightarrow tt$ at the Tevatron \cite{NKAB}; and 
$eu \rightarrow et$ at HERA \cite{NKAB,ABNK}.
As a result,
we show that inclusion of QCD corrections 
significantly stabilizes 
the cross sections.

\section{FCNC top quark cross sections}

We define $s_4=s+t+u-\sum m^2$, with $s,t,u$ standard
kinematical invariants, and where the sum is over the masses squared of 
the particles in the scattering. At threshold $s_4 \rightarrow 0$.
The soft-gluon corrections \cite{KS,NKhq} are of the form 
$[(\ln^l(s_4/m^2))/s_4]_+$, where $l\le 2n-1$
for the order $\alpha_s^n$ corrections.
These corrections are expected to dominate the cross section in the 
near-threshold region, which is relevant for the processes studied here.
The leading logarithms (LL) are those with $l=2n-1$
while the next-to-leading logarithms (NLL) are those with
$l=2n-2$. Here we calculate NLO and NNLO corrections 
in $\alpha_s$ at NLL accuracy, i.e. keeping LL and NLL at
each order in $\alpha_s$. We denote them as NLO-NLL and NNLO-NLL, respectively,
and calculate them using the master formulas in Ref. \cite{NKuni}.

\begin{figure}[!thb]
\vspace*{2.5cm}
\begin{center}
\includegraphics{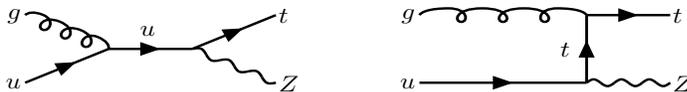}
\caption[*]{Tree-level diagrams for $gu \rightarrow tZ$.}
\label{fig1}
\end{center}
\end{figure}
\vspace{-5mm}

First we study the process  $gu \rightarrow tZ$ 
in $p{\bar p}$ collisions at the Tevatron.
In Fig. \ref{fig1} we show the lowest-order Feynman diagrams.

\begin{figure}[!thb]
\vspace*{7.0cm}
\begin{center}
\includegraphics{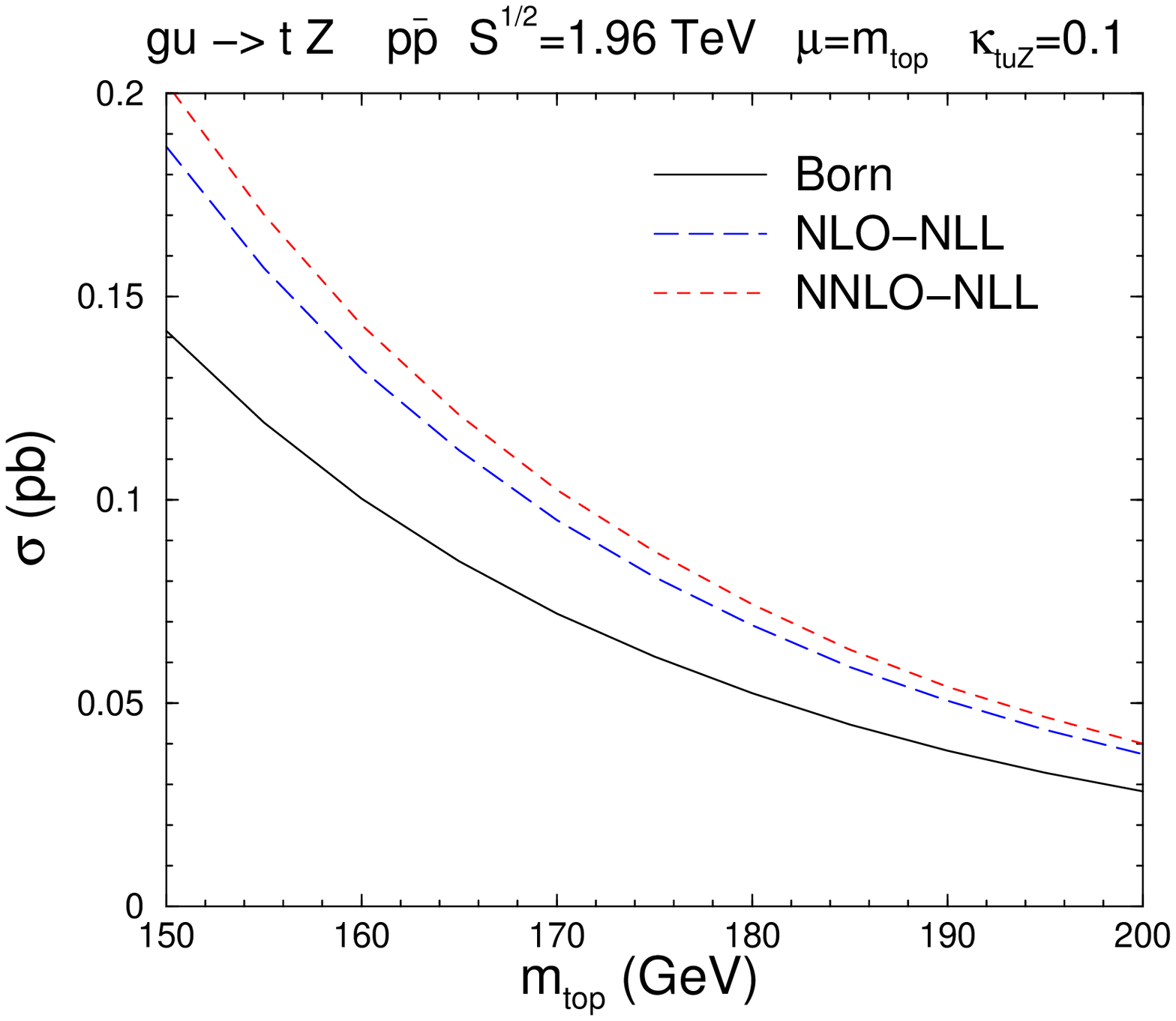}
\includegraphics{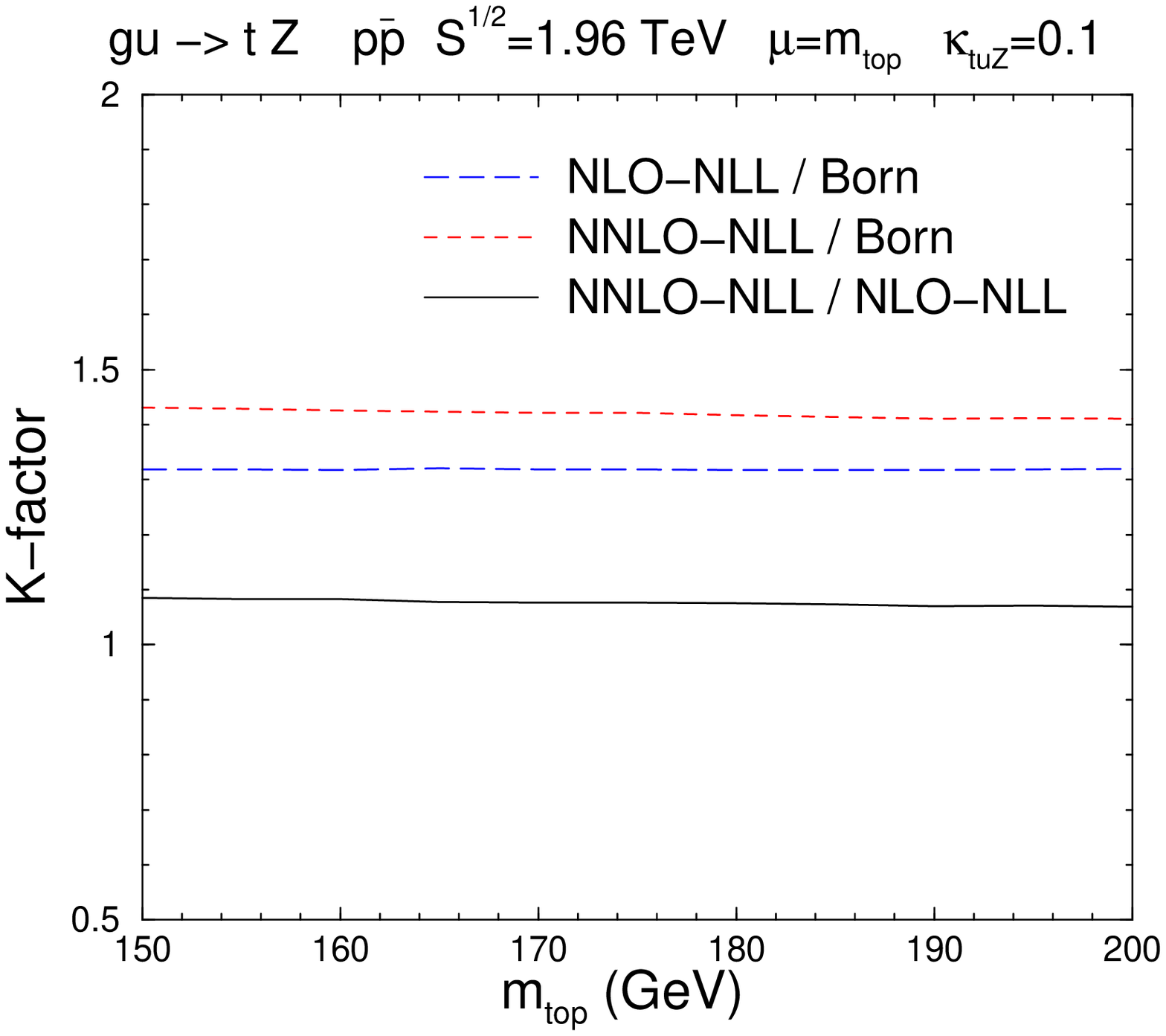}
\vspace{1.5cm}
\caption[*]{Cross sections (left) and $K$-factors (right) 
for $gu\rightarrow tZ$ at the Tevatron.}
\label{fig2}
\end{center}
\end{figure}

In Fig. \ref{fig2} we show plots versus top quark mass 
of the Born, NLO-NLL, and NNLO-NLL
cross sections and of various $K$-factors, which are defined as ratios of 
cross sections at different orders.
Note that $K$-factors are independent of the notation/specification for 
the anomalous couplings. We have set the scale $\mu$ equal to the top quark
mass and set $\kappa_{tuZ}=0.1$. 

\begin{figure}[!thb]
\vspace*{6.0cm}
\begin{center}
\includegraphics{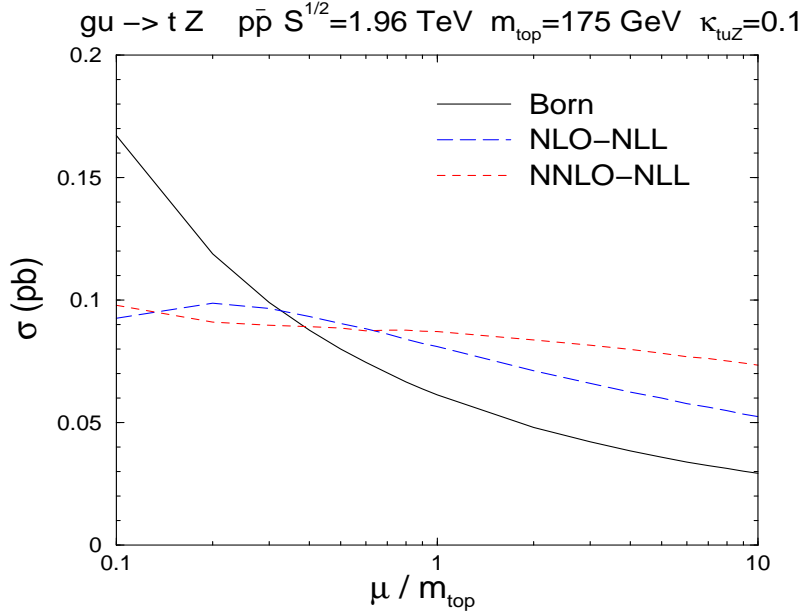}
\vspace{2.0cm}
\caption[*]{The scale dependence of the $gu\rightarrow tZ$ cross section at 
the Tevatron.}
\label{fig3}
\end{center}
\vspace*{-1.cm}
\end{figure}

In Fig. \ref{fig3} we plot the scale dependence of the cross section
for a top mass $m=175$ GeV.
It's clear that the dependence of the cross section on scale
is significantly decreased when we add the NLO-NLL and NNLO-NLL corrections.
For $\mu=m=175$ GeV, $\kappa_{tuZ}=0.1$ and ${\sqrt S}=1.96$ TeV we find
$\sigma_{NNLO-NLL }^{gu\rightarrow tZ} =87^{+2}_{-3}\ {\rm fb}$
where the uncertainty comes from scale variation between $m/2$ and $2m$.
We note that the cross section for the process $gc \rightarrow tZ$,
involving the charm quark, is negligible by comparison.
We also note that the cross section for anti-top production,
$g{\bar u} \rightarrow {\bar t} Z$, is the same 
as for top production.

The results for $gu\rightarrow t\gamma$  are qualitatively the same --
we find again stabilization of the cross section versus scale variation,
as well as a similar cross section level 
($\sigma_{NNLO-NLL }^{gu\rightarrow t\gamma} 
=95^{+17}_{-11}\ {\rm fb}$ for $\mu=m=175$ GeV and $\kappa_{tu\gamma}=0.1$).
In the case of  the process $uu\rightarrow tt$  the cross section is also
stabilized; however, this process is qualitatively different: it has a 
significantly lower cross section
($\sigma_{NNLO-NLL }^{uu\rightarrow tt} =1.74^{+0.00}_{-0.02}\ {\rm fb}$
for $\mu=m=175$ GeV and $\kappa_{tuZ}=\kappa_{tu\gamma}=0.1$)
but a much cleaner signature~\cite{NKAB}.

\begin{figure}[!thb]
\vspace*{2.5cm}
\begin{center}
\includegraphics{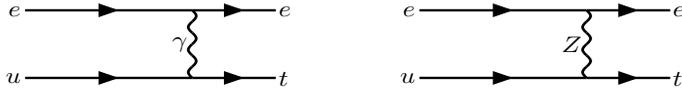}
\caption[*]{Tree-level diagrams for $eu \rightarrow et$.}
\label{fig4}
\end{center}
\vspace*{-1.cm}
\end{figure}

Next we study the process  $eu \rightarrow et$ in $ep$ collisions
at HERA \cite{DIS2000,H1,ZEUS}. 
In Fig. \ref{fig4} we show the lowest-order Feynman diagrams.
\begin{figure}[thb]
\begin{center}
\hspace*{-0.5cm}
\epsfig{width=9cm,height=8.0cm,file=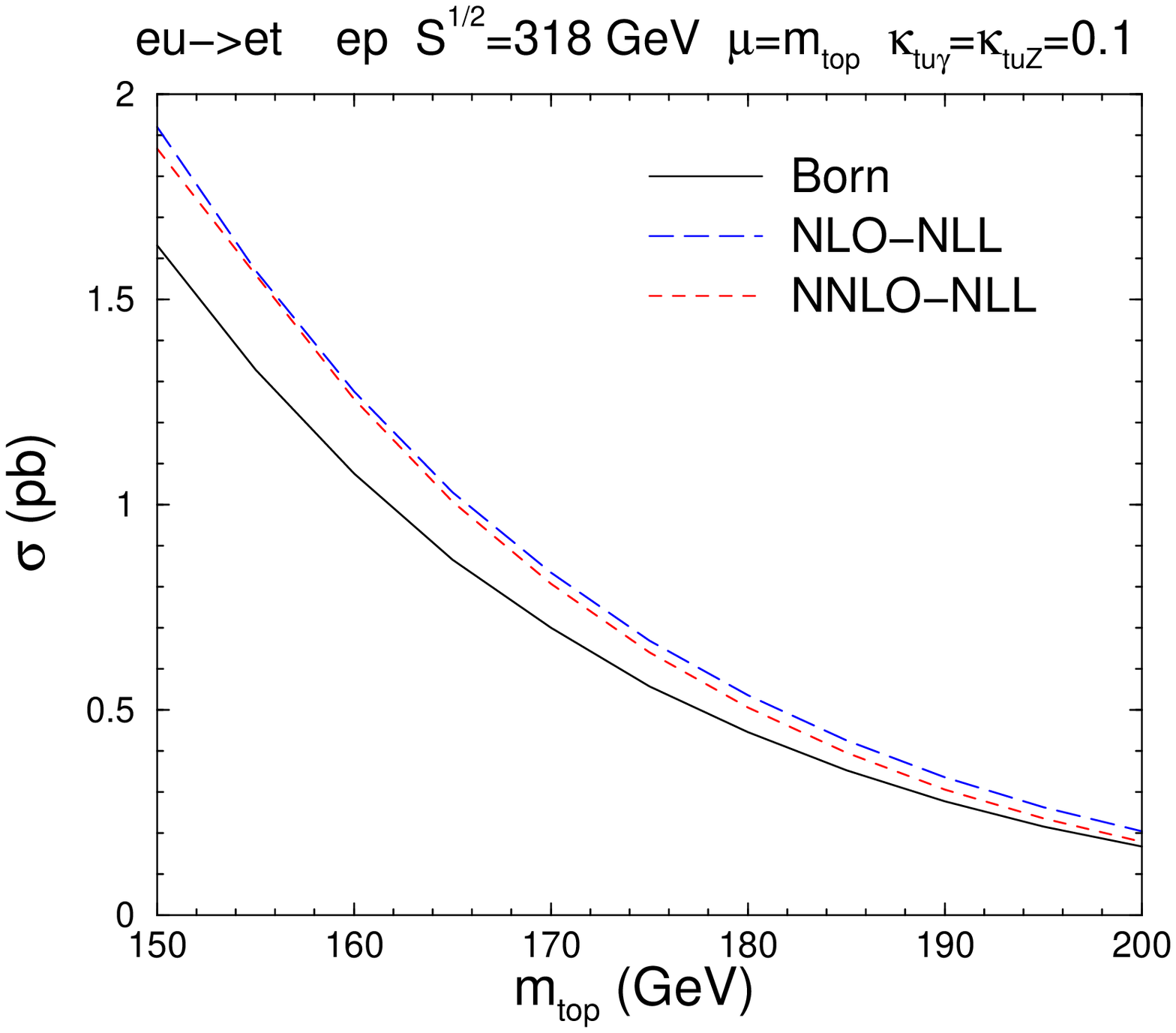}%
\hspace*{-0.5cm}
\epsfig{width=7.5cm,height=8.0cm,file=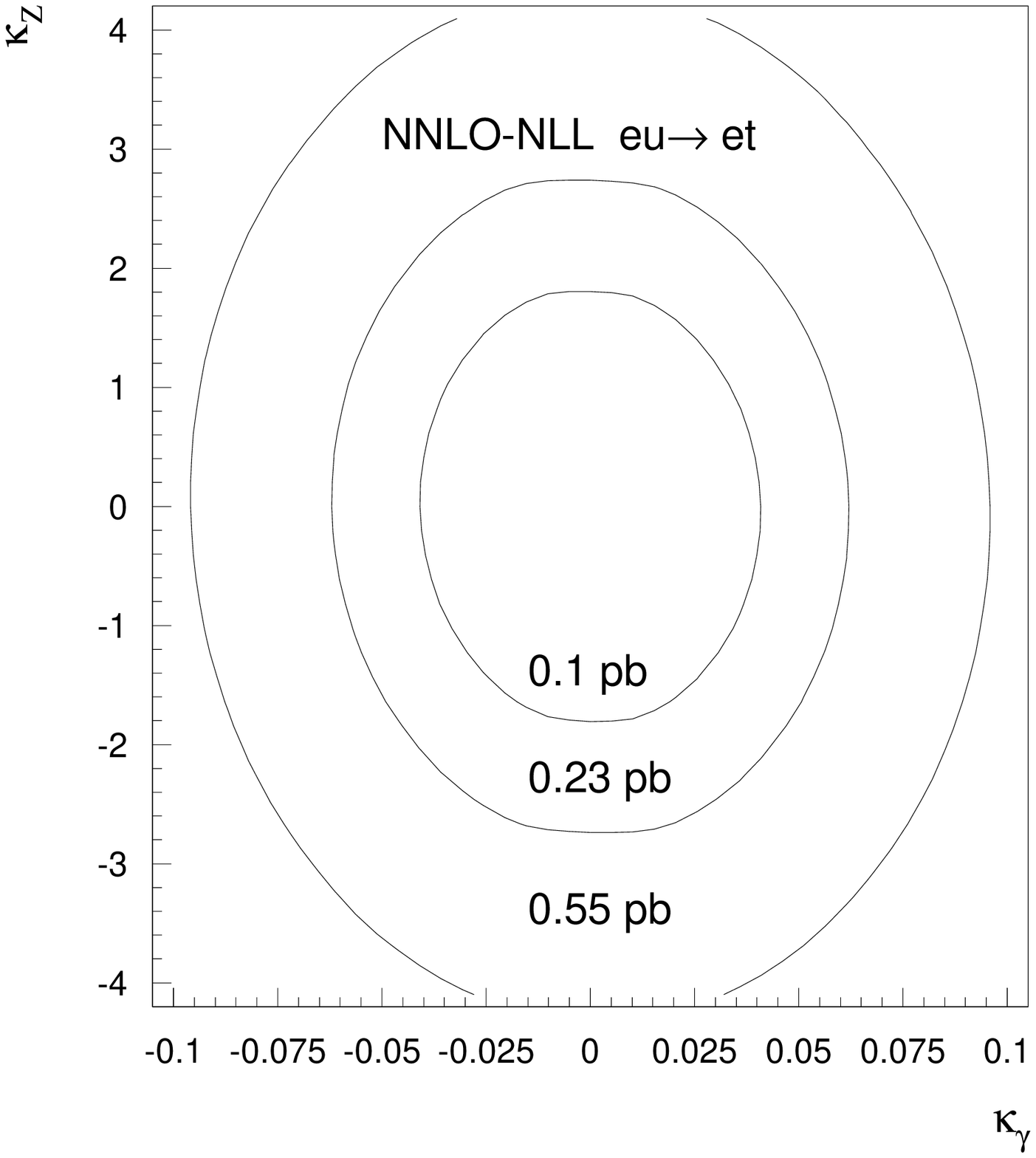}
\caption[*]{Cross sections (left) and HERA reach (right) 
for the process $eu\rightarrow et$.}
\label{fig5}
\end{center}
\end{figure}
In Fig. \ref{fig5} we show plots of the Born, NLO-NLL, and NNLO-NLL
cross sections versus top mass, and of contour levels 
in the $\kappa_{tu\gamma},\kappa_{tuZ}$ plane. We have set $\mu=m$.
It is evident that HERA is much more sensitive to the $\kappa_{tu\gamma}$
coupling than to $\kappa_{tuZ}$.
The NNLO-NLL cross section at HERA for $\mu=m=175$ GeV, 
$\kappa_{tu\gamma}=\kappa_{tuZ}=0.1$ and ${\sqrt S}=318$ GeV is
$\sigma_{NNLO-NLL }^{eu\rightarrow et} =0.64^{+0.05}_{-0.04}\ {\rm pb}$,
where again the uncertainty comes from scale variation between $m/2$ and $2m$.
We note that almost all of the cross section comes from the 
$\kappa_{\gamma}$ coupling.
We also note that contributions from charm are negligible.
In the case of $e{\bar t}$ production, involving the anti-top,
the cross section is quite small
$\sigma_{NNLO-NLL }^{e{\bar u}\rightarrow e{\bar t}}=0.0079$ pb,
and thus assymetrical to $et$ production.

\section*{Acknowledgements} 

The research of N.K. has been 
supported by a Marie Curie Fellowship of the European Community programme 
``Improving Human Research Potential'' under contract no.
HPMF-CT-2001-01221. The work of A.B. was supported in part by the U.S.
Department of Energy.

\end{document}